# Effect of tensile stresses on bainitic isothermal transformation

T. J. Su 12, M. Veaux 1, 3, E. Aeby-Gautier 1, S. Denis 1, V. Brien 1 and P. Archambault 1

1 LSG2M, UMR 7584 du CNRS, INPL, École des Mines, Parc de Saurupt, 54042 Nancy cedex, France
2 Beijing Institute of Technology, Beijing 100081, China
3 PSA, Centre Technique de Belchamp, 25420 Voujeaucourt, France

**Abstract :** The effects of tensile stresses on isothermal bainitic transformation were studied in the case of a 35MV7 steel. The modification of transformation kinetics and the presence of transformation plasticity is shown in a first step. Furthermore the effect of stress on the morphological modifications of the ferrite laths is illustrated. The role of the stress on these changes is analysed.

## 1. INTRODUCTION

It is well established that when transformation occurs in the presence of stresses, the transformation kinetics, the mechanical behaviour as well as the morphology of the product phases are modified [1,2]. If those interactions between stresses and diffusion dependent or martensitic transformations are known and clarified, less results exist in the case of the bainitic transformation. The mechanical behaviour, in term of the transformation plasticity in bainite, as well as the microstructural modifications associated with the transformation under stress are discussed in very few studies [3, 4]. For the prediction of the microstructure formation as well as the calculations of the internal stresses during quenching of steels [5], a good knowledge of the modifications of the bainitic transformation mechanisms is required to understand and further model the interactions of stresses on the phase transformations.

## 2. EXPERIMENTAL PROCEDURES

The influence of an applied tensile stress on the isothermal transformations of a 35MV7 steel was studied. AU tests were performed with our in-house thermomechanical simulator DITHEM which allows to apply simultaneously controlled thermal and mechanical cycles. Radiation furnace or induction heating and controlled cooling by gas (helium) sprayed on the specimen are used. The mechanical load is imposed by an hydraulic jack. The diameter and gage length of the specimen are respectively 3 mm and 20 mm. After austenitization at 1100C during 10 min, the specimen was cooled to the isothermal holding temperature at a rate of 20'ces to 80°C/s, depending on the heating process. The stress was then applied when reaching the transformation temperature (450 or 350C) and maintained until transformation was completed. Applied stresses were ranging from 0 to 235 MPa. TEM observations were performed on some specimens of which the transformation conditions are given in Table 1. At the end of the transformation, the specimen was quenched to room temperature. The time at which transformation was completed is largely dependent on the transformation temperature and the applied stress [6] as shown in Table 1. Thin foils were cut perpendicular to the direction of the tensile stress and examined by transmission electron microscopy (Philips CM200 at 200 kV).

Table 1. Isothermal treatment conditions.

| Sample reference | Transformation Temperature (°C) | Applied tensile stress (MPa) | Transformation Time (s) |
|---|---|---|---|
| 350-0 | 350 | 0 | 500 |
| 350-128 | 350 | 128 | 142 |
| 450-0 | 450 | 0 | 500 |
| 450-144 | 450 | 144 | 142 |

## 3. RESULTS

## 3. 1. Effect of stress on the transformation kinetics and the mechanical behaviour

Figure 1 shows the gage length variations measured during the isothermal transformation under various stresses at 350C. When increasing the stress, an acceleration of the transformation is observed : the incubation period is shortened and the transformation rate increases [1]. Similar results were obtained for the transformation at 450C. Moreover, the transformation plasticity deformation associated with the transformation under stress is clearly evidenced. The evolution of the strain obtained when the transformation is completed, is plotted versus the normalized stress (ratio applied stress / yield stress of the parent austenite) in Fig. 2, for the two transformation temperatures. For a normalized stress less than unity, the results are actually close, whatever the transformation temperature. Moreover, the increase of the transformation plasticity with stress is non linear : the slope $d\varepsilon/d\sigma$ increases for stresses above the yield stress of austenite, especially for the lower transformation temperature. These phenomena are detailed and discussed in [6].

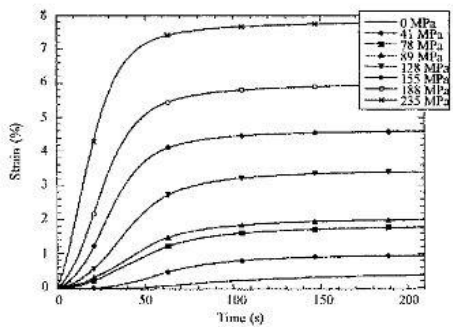

**Figure 1.** Influence of the applied stress on length variations during the bainitic transformation at 350°C.

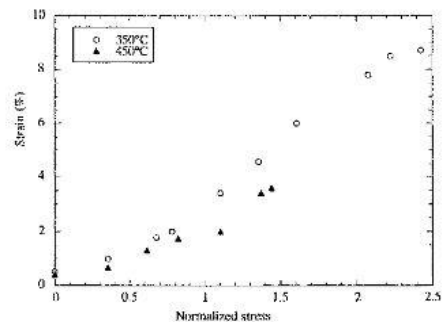

**Figure 2.** Strain variations during the transformation versus normalized applied stress for transformation at 450 and 350°C.

## 3. 2. Analysis of the microstructures

First SEM observations of the microstructure formed at 350C under stress have shown that each austenite grain appears to transform into fewer variants of bainite, giving the microstructure a less random appearance [6]. For the samples transformed under very high stress (normalized stress of 2. 7), each austenite grain has been transformed into a single orientation. These results are consistent with those of Bhadeshia et al. and Matsuzaki et al. [3, 4]. The measurements of the width of the plates formed at different temperatures in the case of the 35MV7 steel showed that there is no large change of the plate width with the increase of the applied stress [6].

In Fig. 3 are reported TEM micrographs of specimens transformed at 350 C with stress (350-128) and without stress (350-0). The microstructure of specimen 350-0 is typical of lower bainite (Fig. 3 (a)). This structure consists of aggregates of platelets and carbides precipitated within these sub-units. To accommodate the internal stress caused by the displacive transformation, several sheaves with different orientations form within a previous austenite grain. The area of a sheave and the size of a platelet can be varied by the section effect. Approximately, the length of the sub-units is about 1 micrometre to 3 micrometres, and the width about 0. 2 micrometres. When transformed under stress, plates grow much longer, with a length of over 10 micrometres and a width between 0. 3 micrometres and 0. 9 micrometres as shown in Fig. 3 (b). Ferrite units exhibit a preferential orientation as compared with those of Fig. 3 (a). This is consistent with previous optical and SEM observations [4, 6].

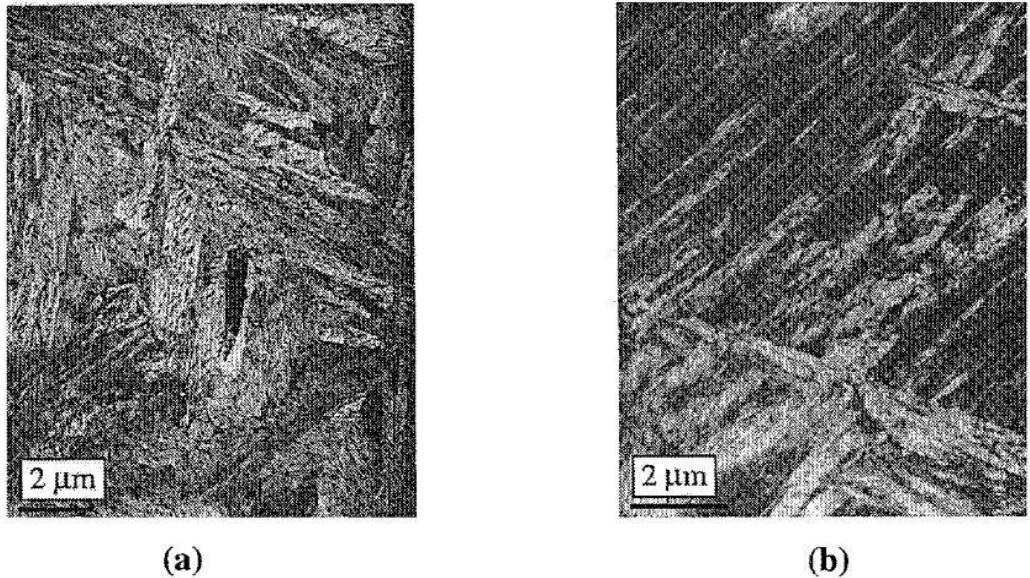

**Figure 3.** TEM micrographs of specimen transformed at 350°C under 0 MPa (a) or 128 MPa (b).

Fig. 4 and Fig. 5 show the TEM micrographs of specimens transformed at 450'C respectively without stress (450-0) and with stress (450-144). Microstructure of 450-0 is typica1 of upper bainite (Fig. 4), consisting of ferrite plates and carbides primarily precipitated between these plates. The length of the sub-units is about 5 gm to 10 gm, and the width of about 0. 3 gm to 0. 9 gm. The tensile stress bas a large effect on the microstructure evolution, as exhibited in Fig. 5. Ferrite units in 450-144 are much coarser and have diverse morphologies. The length of the units increases and can reach 10 micrometres to 20 micrometres, as well as the width between 1 to 2 micrometres. Morphologies of the units can be plate-like (Fig. 5(a)) as without stress can be plate-like (Fig5 (b)) or needle-like.

At last carbides are more difficult to detect. In some areas carbides can be observed as in the upper bainite transformed without stress, while in other places the carbides are no more evidenced. A careful analysis reveals that fine carbides are observed inside the ferrite plates [7].

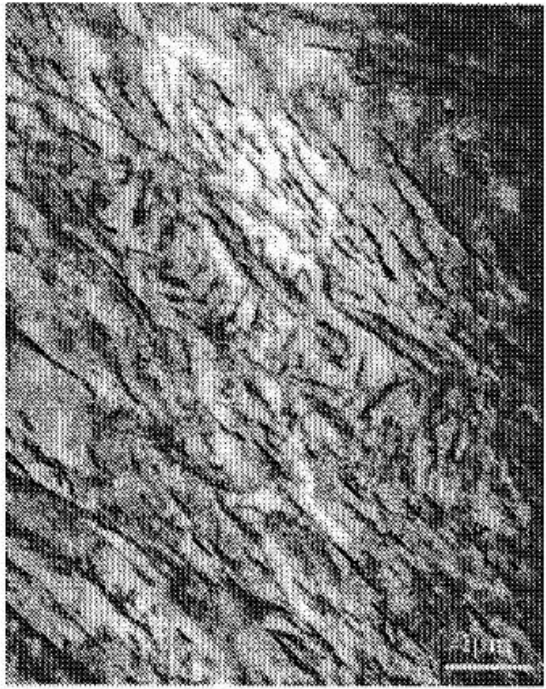

Figure 4. TEM micrograph of specimen transformed at 450°C without stress.

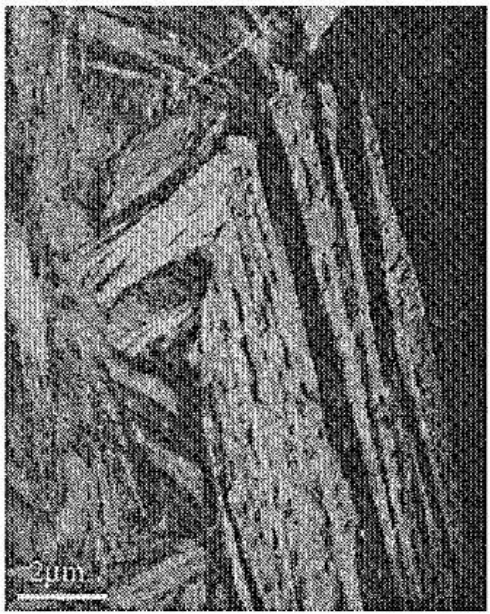

(a)

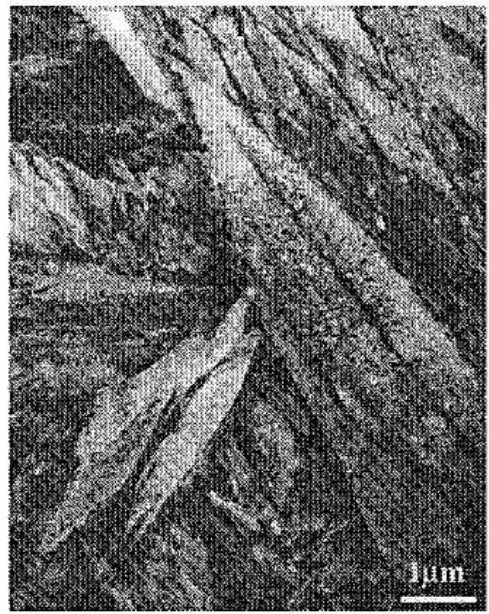

(b)

Figure 5. TEM micrographs of specimen transformed at 450°C under 144 MPa.

## 4. DISCUSSION AND CONCLUSION

The occurrence of transformation plasticity is associated with two basic mechanisms, i. e. the anisotropic plastic slip due to the internal and applied stresses [8], and the preferential orientation of the transformation product by the applied stress [9]. For the transformation conditions reported here, the applied stress is near the yield stress of the parent phase. The observations clearly evidence that a preferential orientation is observed. Thus part of the transformation plasticity is due to ferrite units orientation. Moreover, a careful analysis of the crystallographic orientation along some plates shows that a continuous disorientation exists in the ferrite units, which can be associated with large deformation inside the plates during the transformation [7].

Dislocations were also observed in the parent austenite. This strong evidence of plasticity in both the parent and product phases as well as the preferential morphologic orientation of the ferrite units show that the two transformation plasticity mechanisms have to be considered for the bainitic transformation. The modifications in the morphology, the increase in length and in width of the plates, reveal that the plate's growth is modified. The acceleration of the transformation kinetics can be explained by a modification of the nucleation and growth rates. An increase in the nucleation rate alone, as observed for diffusion controlled transformations [10] or for strain induced martensite [11], would lead to smaller ferrite units. The present observations show that the ferrite sub-units are longer, their growth is thus favored by the stress, rather than their nucleation. As carbides are mainly in the ferrite units when transformation occurs under stress, one can think that the growth rate enhancement is mainly due to the interfacial kinetics modification (easier accommodation of the transformation strain by emission of dislocations) rather than to the increase in the carbon diffusion rate.

**Acknowledgements** The authors are grateful to the support of PSA Peugeot-Citroën and to the French Chinese Program for Scientific Cooperation (PRA 98-MX 98-04).

**References**
[1] : Denis S., Gautier E., Simon A. and Beck G., Mater. Sci. Technol., Vol. 1 (1985) p 805.
[2] : Gautier E., Zhang J. S. and Zhang X. M., J. Phys. IV, Colloque C8, Vol. 5 (1995) p 41.
[3]: Bhadeshia H.K.D.H., David S. A., Vitek J. M. and Reed R. W., Mater. Sci. Technol. Vol. 7 (1991) p686.
[4]: Matsuzaki A., Bhadeshia H. K. D. H. and Harada H., Acta Met. Mater. Vol. 42 (1994) p 1081.
[5] : Denis S., Archambault P., Gautier E., Simon A., Beck G., J. of Materials Engineering and Performance, Vol. 11 (2002) p. 92.
[6] : Veaux M., Louin J. C., Houin J. P., Denis S., Archambault P., Gautier E., J. Phys. IV, France 11 (2001) Pr4-181-188.
[7] : Su T. S., Denis S., Aeby-Gautier E., To be published
[8] : Greenwood G. W. and Johnson R. H., Proc. Roy. Soc., Vol. 283A (1965) p 403.
[9] : Magee C. L., Ph. D. Thesis, Carnegie Mellon University, Pittsburg, PA (1965).
[10] : Gautier E. and Simon A. PTM 87, International Conference Phase Transformations'87 University of Cambridge 6-10 July 1987, The Institute of Metals, Ed. G. W. LORIMER, p 451.
[11]: Oison G.B. and Cohen M. Metall. Trans. A 6A (1975) pp 791-795.